\documentclass[12pt]{article}

\usepackage[dvips]{graphicx}
\usepackage{times}

\begin{document}
\begin{titlepage}

\hfill{LIGO-T970229-R}

\begin{center}

\vfill
{\Large\bf Dynamics of Fabry-Perot Resonators with Suspended Mirrors \\
I. Nonlinear Coupled Oscillators}

\vspace{1cm}
{M. Rakhmanov and A. Arodzero}

\vspace{1cm}
{\it LIGO Project \\
California Institute of Technology \\
Pasadena, CA 91125} 

\end{center}

\vfill
\begin{abstract}
The dynamics of Fabry-Perot cavity with suspended mirrors is described. The
suspended mirrors are nonlinear oscillators interacting with each other
through the laser circulating in the cavity. The degrees of freedom decouple
in normal coordinates, which are the position of the center of mass and the
length of the cavity. We introduce two parameters and study how the dynamics
changes with respect to these parameters. The first parameter specifies how
strong the radiation pressure is. It determines whether the cavity is
multistable or not. The second parameter is the control parameter, which
determines location of the cavity equilibrium states. The  equilibrium state
shows hysteresis if the control parameter varies  within a wide range. We
analyze stability of the equilibrium states and identify the instability
region. The instability is explained in terms of the effective potential: the
stable states correspond to local minima of the effective potential and
unstable states correspond to local maxima. The minima of the effective
potential defines the resonant frequencies for the oscillations of the cavity
length. We find the frequencies, and analyze how to tune them.
Multistability of the cavity with a feedback control system is analyzed in
terms of the servo potential. The results obtained in this paper are general
and apply to all Fabry-Perot cavities with suspended mirrors.
\end{abstract}

\vfill
\end{titlepage}

\newpage
\tableofcontents

\newpage
\section{Introduction}

Very long Fabry-Perot cavities serve as measuring devices for 
interferometric gravitational wave detectors. Several such detectors are
currently under construction \cite{Bradaschia:1990, Abramovici:1992,
Tsubono:1995}. The cavities are planned to have high circulating power and 
large storage time. For example, LIGO (Laser Interferometer Gravitational
wave Observatory) Fabry-Perot cavities will accumulate 10 kW of power and
will have roughly 1 ms of storage time.

The suspended mirrors are designed to move freely along the direction of 
the beam propagation. Due to multi-beam interference in the Fabry-Perot 
cavity the motion of the mirrors strongly affects the light inside the cavity. 
The light, in turn, affects the motion of the mirrors by exerting radiation 
pressure on them. The interaction of light in the cavity with the suspended 
mirrors through radiation pressure gives rise to a nonlinear dynamics. 
Finite time of light propagation in the cavity introduces a time delay in
the dynamics. The propagation time gives rise to storage time. 
Thus a Fabry-Perot cavity is a dynamical system with delay; such
systems are known to have instabilities \cite{Minorsky:1962}.

The significance of the ``spring'' action and the ``damping'' effect of the
radiation pressure for the dynamics of the Fabry-Perot cavity was pointed out
by Braginskii \cite{Braginskii:1977}.  The cavity with one suspended mirror
driven by radiation pressure was studied experimentally by Dorsel et al
\cite{Dorsel:1983, Dorsel:1984, Dorsel:1985, Meystre:1985}. The main results
reported in these papers  are observations of optical bistability and mirror
oscillations with frequencies determined by the radiation pressure. These
authors analyzed  their results using the adiabatic approximation for the
intra-cavity field.  At about the same time Deruelle and Tourrenc studied the
Fabry-Perot  cavity with  suspended mirrors theoretically
\cite{Deruelle:1984, Tourrenc:1985}.  Their analysis  revealed delay-induced
instability caused by the radiation pressure in  the cavity. This instability
was further studied by other researchers  \cite{Aguirregabiria:1987,
Bel:1988}.  Stabilization of the Fabry-Perot cavity by a control system was
discussed by Meers and MacDonald \cite{Meers:1989}. Recently, the radiation
pressure induced dynamics of Fabry-Perot cavities attracted attention of the
VIRGO group in connection with the design of the length control system of
their detector \cite{Chickarmane:1998}.
Similar research has been done in LIGO and is presented in this paper. 

Emerging laser gravitational wave detectors require detailed modeling and
pose new questions for the study of dynamics. From a phenomenological
point of view, there is a question of what are the parameters that define
the universal properties of Fabry-Perot cavities with suspended
mirrors, and how the dynamics changes with respect to these parameters. From
a point of view of applications, there is a question of how to generalize
the results obtained in table-top experiments to large scale Fabry-Perot 
cavities of the gravitational wave detectors. 

In this paper we attempt to provide a phenomenology of the Fabry-Perot 
cavities for modeling and optimization of the performance of LIGO 
interferometers. Due to the complexity
of the subject we split the discussion into two papers. In the first paper
we study various aspects of the nonlinearity in the dynamics, leaving aside
the time delay instabilities. In the second paper \cite{Rakhmanov:1998b} we
consider the time delay instabilities and study the dynamics of a 
Fabry-Perot cavity with a realistic control system. 

In this paper we formulate the dynamics in terms of normal coordinates: 
the cavity length and the cavity center of mass. We show that a small 
part of the radiation pressure in the cavity, the excess force, excites 
the cavity center of mass. In absence of the excess radiation pressure, the 
dynamics of the cavity length is equivalent to the dynamics of a suspended 
mirror in a cavity, which has one mirror suspended and one mirror fixed. 
To study the universal properties of the cavity dynamics, such as 
multistability, we introduce two parameters. The
first parameter is a control parameter which allows us to select
the equilibrium state. The second parameter characterizes strength of the 
radiation pressure and determines 
whether the system is stable or multistable and 
how many equilibrium states are there. 
The results obtained in this paper are general and apply
to any Fabry-Perot cavity with suspended mirrors. Numerical calculations and
modeling with parameters of LIGO cavities are given throughout this paper.

The paper is organized as follows. In Section 2 we describe the equations of
motion and define the control parameter. In Section 3 we formulate the 
dynamics in normal coordinates. In Section 4 we construct equilibrium states 
and introduce the specific radiation pressure. In Section 5
hysteresis and instabilities are described. In Section 6 the 
global properties of the dynamics are analyzed in terms of the
effective potential. In Section 7 we provide explicit formulas and numerical
values for the cavity resonant frequencies. In Section 8 we briefly describe 
nonlinear dynamics of the cavity with a control system.

\newpage
\section{Dynamical Equations and Regimes}

\subsection{Mirror Coordinates}

\begin{figure}[ht]
\begin{center}
  \caption{Coordinates of mirrors and their suspension points}
  \label{coords}
  \includegraphics[width=12cm]{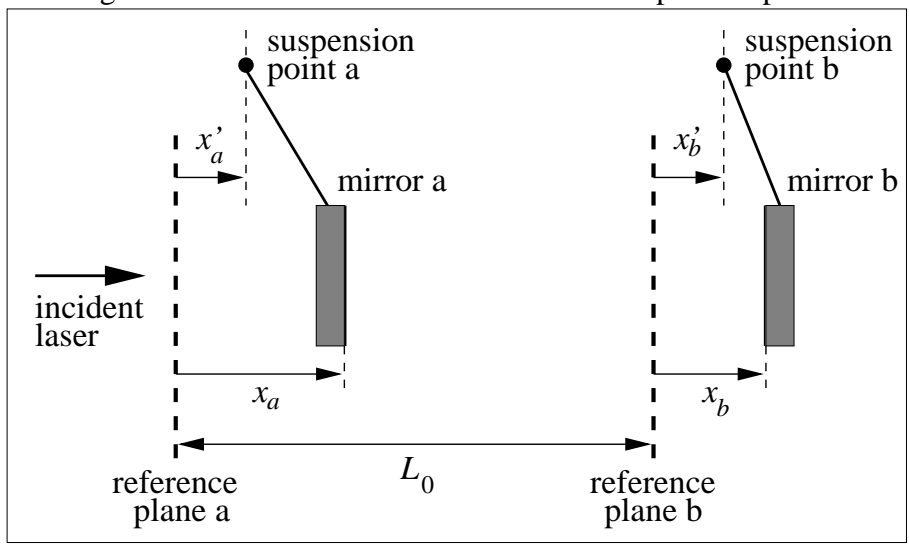}
\end{center}
\end{figure}

The dynamical system we study in this paper is a Fabry-Perot 
cavity with two suspended mirrors, labeled by $a$ and $b$, 
and a laser incident on the cavity from one side. 
A suspended mirror is a rigid body with six degrees of freedom, 
whose dynamics depends on the suspension design. 
In this paper we neglect angular degrees of freedom  of the 
mirrors and consider only the motion of their center of mass. 
We also neglect the side motion of the mirrors. In this approximation 
the mirrors are always aligned to the incident beam 
and move parallel to themselves as simple pendula. We specify the 
mirror positions by their coordinates, $x_a(t)$ and 
$x_b(t)$, in the inertial coordinate frames
as shown on figure~\ref{coords}. The mirror suspension points are 
also moving, their coordinates are $x'_a(t)$ and 
$x'_b(t)$. 

The two inertial coordinate frames are separated by a large distance, 
the nominal cavity length, $L_0$, which we consider fixed. The actual length 
of the cavity depends on mirror positions in their respective coordinate 
frames:
\begin{equation}\label{Linst}
   L = L_0 + x_b(t) - x_a(t). 
\end{equation}
The large cavity length of the gravitational wave detectors makes the  
delay time,
\begin{equation}
   T = \frac{L_0}{c},
\end{equation}
to be non-negligible. For example, in LIGO interferometers the cavities are 
4km long and the corresponding delay time is roughly 13 $\mu$s.

\subsection{Equations for Field Dynamics}

Multibeam interference of laser in long Fabry-Perot cavities of 
gravitational wave detectors is far more complex than a similar process 
in short Fabry-Perot interferometers with fixed length 
which are commonly used for laser 
spectroscopy. Both the motion of suspended mirrors and the delay times 
affect the interference and give rise to a complex dynamics. 
We describe dynamics of the laser in the cavity in 
terms of complex amplitudes of the electric field. These complex 
amplitudes correspond to the traveling electro-magnetic waves and 
are defined at some locations. Usually the amplitudes are defined 
at the mirror surfaces. However, the suspended mirrors are 
moving and such a choice would require account for Doppler effects. 
Instead, we define the amplitudes at the reference planes which 
are fixed in the inertial coordinate frames. 

\begin{figure}[ht]
\begin{center}
  \caption{Notations for fields and reference planes}
  \label{fields}
  \includegraphics[width=8cm]{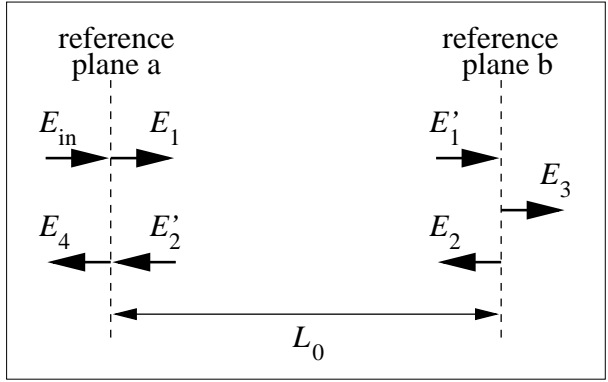}
\end{center}
\end{figure}

Our notations for the amplitudes of the electro-magnetic waves are shown 
in Fig.~\ref{fields}. 
The amplitude of the wave, incident on the front mirror, is $E_{in}$. 
The amplitudes of the forward-propagating wave inside the cavity, 
defined at the two reference planes, are $E_1$ and $E'_1$. 
Similarly, the amplitudes of the backward-propagating wave are 
$E_2$ and $E'_2$. 
The amplitudes of the cavity transmitted and reflected waves are $E_3$ 
and $E_4$.

Propagation of the laser inside the cavity is described by the delay 
equations: 
\begin{eqnarray}
   E'_1(t) & = & E_1(t - T), \label{Ep1} \\
   E'_2(t) & = & E_2(t - T). \label{Ep2}
\end{eqnarray}
These and other equations for the complex amplitudes are derived in 
the Appendix~\ref{defE}. The equations for reflection and transmission 
of the laser at the end mirror are: 
\begin{eqnarray}
   E_2(t) & = & - r_b e^{- 2ik x_b(t)} E'_1(t), \label{E2} \\
   E_3(t) & = & t_b E'_1(t), \label{E3} 
\end{eqnarray}
where $r_b$ is the reflectivity and $t_b$ is the transmissivity of the 
end mirror.

Using the superposition principle we obtain similar equations for the 
front mirror: 
\begin{eqnarray}
   E_1(t) & = & t_a E_{in}  - r_a e^{ 2ik x_a(t)} E'_2(t), \label{E1} \\
   E_4(t) & = & t_a E'_2(t) + r_a e^{-2ik x_a(t)} E_{in}, \label{E4}
\end{eqnarray}
where $r_a$ and $t_a$ is the reflectivity and the transmissivity of the 
front mirror. Numerical values for the reflectivities and the trasmissivities 
of the mirrors in LIGO cavities are given in the Appendix~\ref{cavPar}.

The equations (\ref{Ep1}-\ref{E4}) describe dynamics of the laser interacting 
with the moving mirrors of the cavity. These equations can be reduced to 
one equation with one field, which is called the self-consistent field. 
All other fields 
can be expressed in terms of the self-consistent field through the 
field equations above. 

We choose the forward-propagating field to be the self-consistent field. Let 
the amplitude of the self-consistent field be $E(t)$ (same as $E_1(t)$). 
Then the equation for dynamics of the self-consistent field is 
\begin{equation}\label{selfCon}
   E(t) = t_a E_{in} + r_a r_b e^{-2ik z(t)} E(t - 2T),
\end{equation}
where $z(t)$ is the relative mirror position defined as 
\begin{equation}\label{xdef}
   z(t) = x_b(t - T) - x_a(t).
\end{equation}
The quantity, $L_0 + z(t)$, is the distance between the 
mirrors in relativistic sense. 

Since the speed of light is finite we cannot obtain 
information about the instantaneous cavity length defined in 
eq.~(\ref{Linst}). 
However, we can measure the relativistic cavity length, $z(t)$, by  
observation of one of the cavity fields. This can be done using 
the Pound-Drever signal extraction scheme. The Pound-Drever signal 
is based on observation of the cavity reflected laser
\begin{eqnarray}\label{VPD}
   V_{\mathrm{P-D}} & \sim & \mathrm{Im} \{ e^{2ikx_a(t)} E_4(t) \} \\
   & \sim & \mathrm{Im} \{ E(t) \},
\end{eqnarray}
which is defined in terms of the self-consistent field and, therefore, 
depends on the cavity length, $z(t)$. 

In the expression, eq.~(\ref{xdef}), the delay appears in the coordinate 
of the end and not the front mirror. 
This is because upon entering the cavity the laser 
first reflects off the end mirror and then reflects 
off the front mirror.


\subsection{Forces Acting on Mirrors}

Dynamics of the suspended mirrors is defined by the forces acting on them. 
Since we analyze only longitudinal motion of the mirrors we consider 
only horizontal components of the forces. 
The horizontal component of the wire tension, acting on the front mirror, is
\begin{equation}
   - m_a \omega_0^2 (x_a - x'_a),
\end{equation}
where $m_a$ is the mass of the front mirror and 
$\omega_0$ is the pendulum frequency. There is a similar expression 
for the end mirror. 

The laser circulating in the cavity and the incident laser 
exert pressure on the 
mirrors. Let the radiation pressure on the front and the end mirror 
be $R_a$ and $R_b$. These forces are defined by the amplitudes of 
the fields as follows:
\begin{eqnarray}
   R_a & = & \frac{1}{c} \left\{ |E_{in}|^2 - |E_1|^2 
                - |E'_2|^2 + |E_4|^2 \right\}, \label{Ra} \\
   R_b & = & \frac{1}{c} \left\{ |E'_1|^2 + |E_2|^2 
                - |E_3|^2 \right\}. \label{Rb}
\end{eqnarray}

Many control schemes, including the control system of LIGO arm cavities, 
require actuators on the mirrors. The actuators can be coil-magnet pairs, 
electro-static boards or other devices. For our analysis the details 
of the actuators are not important. We only assume that actuators are 
perfectly aligned and apply forces 
to the mirrors in the direction of the beam propagation. 
Let the actuator forces be $F_a^{\mathrm{act}}$ and $F_b^{\mathrm{act}}$. 

There is also a friction force due to energy 
losses in the suspension wires and the actuators. 
We assume that the friction is proportional to velocity. Then the friction 
force acting on the front mirror can be written as
\begin{equation}
   - m_a \gamma \dot{x}_a,
\end{equation}
where $\gamma$ is the damping coefficient. A similar expression can be 
obtained for the end mirror. Typically the damping coefficient 
is very small: $\gamma/\omega_0 \sim 10^{-5}$. 

Numerical values for the parameters 
of LIGO interferometers are given in Appendix~\ref{cavPar}.

\subsection{Control Parameter}

Let us consider the sum of the pendulum restoring force and the actuator force 
for the front mirror:
\begin{equation}\label{comb}
   - m_a \omega_0^2 (x_a - x'_a) + F_a^{\mathrm{act}}. 
\end{equation}
There are two parameters here: the position of the suspension 
point and the actuator force. The mirror can be controlled 
through both the actuator and the suspension point. 
However, the dynamics depends only on the sum of the two parameters, 
and not separately on each one. 
Therefore, we combine the two parameters into a new parameter 
\begin{equation}\label{controlPar}
   u_a = x'_a + \frac{F_a^{\mathrm{act}}}{m_a \omega_0^2}.
\end{equation}
This is the control parameter for the front mirror. 
A similar parameter, $u_b$, can be introduced for the end mirror. 

The control parameter has dimension of length and the meaning of an equivalent 
displacement of the suspension point. The equivalence is based on the 
conversion factor $(m \omega_0^2)^{-1}$, 
which for LIGO is roughly 4.2 millimeter per newton of 
the actuator force. 

Moving the suspended mirrors at high frequencies is easier using the actuator.  
However, applying large displacements to the mirror at low frequencies 
is better through the suspension point. Therefore in a 
realistic control system the control parameter, $u$, may obtain its high 
frequency component from the actuator force and its low frequency component 
from the position of the suspension point.

\subsection{Equations of Motion for Mirrors}

The equations of motion for the mirrors are 
\begin{eqnarray}
   \ddot{x}_a + \gamma \dot{x}_a + \omega_0^2 x_a 
   & = & \omega_0^2 u_a + \frac{R_a}{m_a},\\
   \ddot{x}_b + \gamma \dot{x}_b + \omega_0^2 x_b 
   & = & \omega_0^2 u_b + \frac{R_b}{m_b}.
\end{eqnarray}
These are the equations for a pair of damped harmonic oscillators 
driven by forces of the radiation pressure. The oscillators 
are not independent: they are interacting with each other through the 
laser which acts as a spring connecting the mirrors. 
The ``spring'' is nonlinear and has time delay.

\subsection{Dynamical Regimes}

A Fabry-Perot cavity has two intrinsic time scales: one is the delay time, 
$T$, the other is the storage time  
\begin{equation}
   \tau = \frac{2T}{|\ln (r_a r_b)|} .
\end{equation}
Correspondingly, there are two intrinsic frequencies defined by the 
cavity: the free spectral range ($f_{\mathrm{FSR}} = (2T)^{-1}$) 
and the cavity low pass frequency 
\begin{equation}
   f_{cav}= \frac{1}{2\pi \tau} .
\end{equation}
Therefore, there are three dynamical regimes for the Fabry-Perot cavity. 

Let $\delta t$ be a characteristic time for a dynamic process 
involving changes in the mirror positions and fields. Let $f$ be the 
characteristic frequencies of the process. The three regimes are:

\begin{enumerate}
   \item decoupled regime: $\delta t \ll T$  ($f \gg f_{\mathrm{FSR}}$), 
   \item delay regime:     $T \ll \delta t \ll \tau$ 
                              ($f_{cav} \ll f \ll f_{\mathrm{FSR}}$),
   \item adiabatic regime: $\delta t \gg \tau$ ($f \gg f_{cav}$).
\end{enumerate}

The first regime corresponds to very 
fast changes in mirror positions and fields in the Fabry-Perot cavity. 
The decoupled regime can be understood on the following example. 
First the mirrors are at rest. Then one mirror 
acquires some velocity, changes its position and stops. If the entire 
process takes time less than $T$ 
the second mirror becomes affected by the process 
after the first mirror comes to rest. 
Therefore, during such process the mirrors are independent of each other.
This is nothing but a relativistic causality limit. In LIGO interferometers 
the processes with frequencies above 37.5 kHz belong to the decoupled 
regime. 

The second regime takes place when the changes in the mirror position are  
faster than transient processes in the cavity but not fast enough to 
decouple the mirrors. In this regime the transients caused by the 
mirror motion never die and the cavity never reaches equilibrium. 
In such a regime the 
motion of one mirror is constantly affected by the other mirror and 
the delays play crucial role. Dynamic processses in LIGO cavities 
with frequencies between 100 Hz and 10 kHz belong to this regime.

The third regime takes place when the mirrors 
move so slow that the laser transients can be neglected. In this case 
we can think of the laser circulating in the cavity as being in 
equilibrium with the mirrors. This is the adiabatic regime. In this 
regime the delays are not important. The adiabatic regime takes place 
in LIGO cavities at frequencies less than 100 Hz.

\subsection{Self-consistent Field in Adiabatic Regime}

In the adiabatic regime the mirror motion is so slow that we can neglect 
the delay time in the equation for the self-consistent field 
eq.~(\ref{selfCon}) and obtain a solution:
\begin{equation}\label{Eadiab}
   E(t) = \frac{t_a \; E_{in}}{1 - r_a r_b e^{-2ik z(t)}}.
\end{equation}
The value of the field in the adiabatic regime is completely defined by 
the current value of the cavity length, $z(t)$. 
The power in the cavity, which we define as the momentum carried by 
the forward-propagating wave ($P = |E|^2$) is also a function of the cavity 
length. The maximum power, $P_{max}$, depends on the mirror parameters 
and defines the cavity gain: 
\begin{equation}\label{Gcav}
  G_{cav} \equiv \frac{P_{max}}{P_{in}} = \frac{t_a^2}{(1 - r_a r_b)^2} .
\end{equation}
The power as a function of length is described by the Airy formula 
\begin{equation}\label{Padiab}
   P = \frac{P_{max}}{1 + F \sin^2 k z(t)} ,
\end{equation}
also known as Airy intensity profile (see \cite{Born:1980}). 
The constant, $F$, is the coefficient of finesse 
\begin{equation}
   F = \frac{4 r_a r_b}{(1 - r_a r_b)^2} .
\end{equation}
The relation between the parameter, $F$, and the finesse of Fabry-Perot 
cavity is given in Appendix~\ref{Airy}.

\newpage
\section{Normal Coordinates}

\subsection{Forces due to Radiation Pressure}

The suspended mirrors are interacting with each other 
through the laser in the cavity. Therefore, the degrees of freedom 
which correspond to the mirrors are coupled. However, we can 
formulate the dynamics in terms of new degrees of freedom which are 
largely independent of each other.  These new degrees of freedom are 
similar to the normal coordinates usually appeared in theory of 
coupled linear oscillators. 
Before we formulate the dynamics in terms of the normal coordinates 
we reduce the number of fields entering the general expressions 
for the radiation pressure on the mirrors, equations~(\ref{Ra}-\ref{Rb}).

The radiation pressure on the end mirror, eq.~(\ref{Rb})  
can be expressed entirely in terms of the self-consistent field: 
\begin{equation}\label{Rb2}
   R_b(t) = \frac{1}{c} (1 + r_b^2 - t_b^2) |E(t - T)|^2 .
\end{equation}
The radiation pressure on the front mirror is produced by the laser  
circulating in the cavity and the laser incident on the cavity, 
eq.~(\ref{Ra}). 
A constant flux of the incident laser produces constant 
radiation pressure on the front mirror: 
\begin{equation}\label{Rin}
   R_{in} = \frac{1}{c} (1 + r_a^2 - t_a^2) |E_{in}|^2 . 
\end{equation}
Therefore, we can write the 
radiation pressure on the front mirror as the sum: 
\begin{equation}
   R_a(t) = - R_b(t - T) + R_{in} + \Delta R(t) .
\end{equation}
In this equation the first term implies that the same force which is  
pushing the end mirror now will be pushing the front mirror in the opposite 
direction after time $T$. The second term is proportional to the power of 
the incident laser and accounts for the constant radiation pressure 
from outside the cavity. The last term, $\Delta R(t)$, which can be 
called the excess radiation pressure, 
accounts for the asymmetry between the mirrors and the interference between 
the incident laser and the laser circulating in the cavity. 
The excess force can also be expressed entirely in terms of the 
self-consistent field and the cavity length:
\begin{equation}\label{DR}
   \Delta R(t) = \frac{\alpha}{c} |E(t - 2T)|^2 -  
               \frac{4}{c} \; t_a r_a r_b \; \mathrm{Re}
               \{ e^{-2ik z(t)} E^*_{in} E(t - 2T) \},
\end{equation}
where $\alpha = 1 - t_b^2 - r_b^2(r_a^2 - t_a^2)$, and 
the asterisk stands for complex conjugation. 

In the adiabatic regime the excess force can be written explicitly as 
a function of length: 
\begin{equation}
   \Delta R = c^{-1} P_{max}
      \frac{\alpha + 4 r_a r_b (r_a r_b - \cos 2kz)}{1 + F \sin^2 kz}.
\end{equation}
Numerical estimates with LIGO parameters show that the excess radiation 
pressure, $\Delta R$, is roughly four orders of magnitude less than the 
radiation pressure, $R_b$. 

The constant radiation pressure, eq.~(\ref{Rin}), 
has no effect on the dynamics, but 
the excess force, eq.~(\ref{DR}), makes the radiation pressure on the mirrors 
unbalanced and affects the dynamics.

\subsection{Normal Modes}

The normal coordinates are the position of the center of mass and the 
cavity length: 
\begin{eqnarray}
   Z(t) & = & \frac{m_a x_a(t) + m_b x_b(t - T)}{m_a + m_b},\\
   z(t) & = & x_b(t - T) - x_a(t).
\end{eqnarray}
Note that both normal coordinates are affected by the delay due to the 
finite time of laser propagation in the cavity. 

Similarly, we introduce the control parameters for the new coordinates:
\begin{eqnarray}
   U(t) & = & \frac{m_a u_a(t) + m_b u_b(t - T)}{m_a + m_b} + 
              \frac{R_{in}}{c (m_a + m_b) \omega_0^2}, \\
   u(t) & = & u_b(t - T) - u_a(t) - \frac{R_{in}}{c m_a \omega_0^2} .
\end{eqnarray}
To make the subsequent formulas simpler we hid the constant 
radiation pressure into the new control parameters. Also we define 
the reduced mass in a usual way:
\begin{equation}\label{rMass}
   \mu = \frac{m_a m_b}{m_a + m_b} .
\end{equation}

In normal coordinates the dynamical equations are the equation for 
the center of mass and the equation for the cavity length: 
\begin{eqnarray}
   \ddot{Z} + \gamma \dot{Z} + \omega_0^2 Z & = & 
      \omega_0^2 U + \frac{\Delta R}{m_a + m_b},\\
   \ddot{z} + \gamma \dot{z} + \omega_0^2 z & = & 
      \omega_0^2 u + \frac{R_b}{\mu} - \frac{\Delta R}{m_a}. 
\end{eqnarray}
We see that the 
radiation pressure mostly affects one degree of freedom: the cavity 
length, $z(t)$. This is because the pressure on the front mirror 
is roughly equal and opposite to the pressure on the end mirror 
if we neglect the constant pressure from outside. 
The motion of the center of mass is affected by the radiation pressure 
only through the excess force.

\subsection{Center of Mass and Observability}

Dynamics of the center of mass of free cavity is defined exclusively 
by the excess radiation pressure, which, in turn, is defined by 
the cavity length, $z(t)$. As a result, changes in the cavity length 
excite motion of the center of mass. On the 
contrary, arbitrary motion of center of mass has no effect on the cavity 
length. Therefore we can ignore the dynamics of the center of 
mass and focus on the cavity length. However, few important points must be 
taken into account before we abandon the motion of the center of mass. 

Although small motion of the center of mass ($\sim 1\mu$m) 
has no effect on the detector performance, large motion ($\sim$ 0.1 mm) 
will cause variations of the actuator gains. 
Therefore, the motion of the cavity center of mass 
must be suppressed to the extent defined by the actuator's range. 
In particular, we must avoid exciting the motion of the center 
of mass by a control system. In gravitational wave detectors the 
control system suppresses motion of the mirror under seismic disturbances 
by applying forces to the mirrors through the actuators. 
Therefore the control system can excite the motion of the 
center of mass, unless the control signals are synchronized:
\begin{equation}
   \delta u_a(t) = - \frac{m_b}{m_a} \; \delta u_b(t - T).
\end{equation}

Unlike the cavity length the center of mass is not 
observable in a strict relativistic sense. To obtain information about 
the cavity's center of mass requires setting up an inertial frame, 
which is impossible for the ground based gravitational wave detectors. 
This poses a problem of observability for control of the 
Fabry-Perot cavity, which can be stated as follows. The cavity length 
is known with high precision (better than $10^{-13}$m) but the center of 
mass can only be measured approximately (up to 1$\mu$m). 
Therefore the cavity length can only be controlled 
approximately. Since the two degrees of freedom are largely independent 
poor observability and control of the center of mass do not affect 
the dynamics of the cavity length. Therefore, 
as long as the motion of the center of mass remains small, it can be ignored.

\subsection{Cavity Length}

The degree of freedom most important for understanding the response of
Fabry-Perot cavity as a part of the  gravitational wave detector is the
cavity length $z(t)$.

In the following discussion of the cavity length dynamics we 
neglect the excess radiation pressure, $\Delta R$. 
To simplify following formulas we make the approximation: 
$1 + r_b^2 - t_b^2 \approx 2$ in the eq.~\ref{Rb2} and write 
the radiation pressure on the end mirror as 
\begin{equation}
   R_b(t) = \frac{2}{c} |E(t - T)|^2 .
\end{equation}
This is a good approximation for high reflective mirrors, such as 
the end mirrors of LIGO cavities.

The dynamics of the cavity length is independent of the dynamics of the 
center of mass and can be studied separately.
The equations for the cavity length dynamics include only the 
cavity length and the self-consistent field:
\begin{eqnarray}
   & & \ddot{z} + \gamma \dot{z} + \omega_0^2 z = 
       \omega_0^2 u + \frac{2}{c \mu} |E(t-2T)|^2 , 
       \label{pendEq} \\
   & & E(t) = t_a \; E_{in} + r_a r_b \; e^{-2ik z(t)} \; E(t - 2T) 
       \label{iterEq},
\end{eqnarray}
The iteration equation~(\ref{iterEq}) allows us to calculate the amplitude
$E(t)$ at discrete times. The time step is the cavity round-trip time, $2T$.
The value of the amplitude between the steps can be found by 
delaying its value at the last iteration. 

These equations are equivalent to the equations of the 
dynamics of the Fabry-Perot cavity with front mirror fixed and the end mirror 
suspended from wires, provided the mass of the end mirror is equal 
to the reduced mass.

\newpage
\section{Equilibrium States}

\subsection{Static Solutions}

Equilibrium states of the cavity are static solutions of the 
equations (\ref{pendEq}) and (\ref{iterEq}). For static states 
these equations simplify and can be written as 
\begin{eqnarray}
   & & \mu \omega_0^2 (z - u) = \frac{2}{c} \; |E|^2 , \label{equilX}\\
   & & E = \frac{t_a E_{in}}{1 - r_a r_b e^{-2ikz}}. \label{equilE}
\end{eqnarray}
The complex field, $E$, can be eliminated from the 
equations~(\ref{equilX}) and (\ref{equilE}) and the equilibrium condition 
can be written entirely in terms of the cavity length:
\begin{equation}\label{equilx}
   \mu \omega_0^2 (z - u) = \frac{R_{max}}{1 + F \sin^2 kz}.
\end{equation}
Here $R_{max}$ is the maximum radiation pressure on the end mirror 
which corresponds to the maximum power in the cavity:
\begin{equation}\label{Rmax}
   R_{max} = \frac{2}{c} P_{max}.
\end{equation}
The equation~(\ref{equilx}) shows that in equilibrium the radiation pressure 
is balanced by the pendulum restoring force. 

Numerical values of the power and the radiation pressure in LIGO 
Fabry-Perot cavities are given in the Table \ref{mags}. 
\begin{table}[ht]
   \caption{Power and radiation pressure in LIGO Fabry-Perot cavities}
   \label{mags}
   \begin{center}
   \begin{tabular}{lccc}
   \hline
   \hline
                        & $P_{in}(W)$
                        & $P_{max}(W)$ 
                        & $F_{max}(N)$ \\ 
   \hline
   40m LIGO Prototype   & $0.25$
                        & $160$
                        & $1.1 \times 10^{-6}$ \\
   4km LIGO             & $90$
                        & $12000$ 
                        & $7.8 \times 10^{-5}$ \\ 
   \hline
   \hline   
   \end{tabular}
   \end{center}
\end{table}

\subsection{Strength of Radiation Pressure}

To characterize strength of the radiation pressure we introduce a 
dimensionless parameter 
\begin{equation}
   \sigma = \frac{k R_{max}}{\mu \omega_0^2} ,
\end{equation}
where $k = 2\pi/\lambda$ is the wave number, $\mu$ is the reduced mass
and $\omega_0$ is the pendulum frequency. 
The parameter $\sigma$ shows how strong the radiation pressure 
is compared to the pendulum restoring force.
This parameter allows us to characterize nonlinearities induced by the 
radiation pressure quantitatively. As a measure of 
nonlinearities, the parameter $\sigma$ is similar to the Reynolds number 
in fluid dynamics. 

The parameter $\sigma$ also allows us to establish equivalence of the 
equilibrium states of different Fabry-Perot cavities. Namely, the 
equilibrium states of a Fabry-Perot cavity with high 
circulating power and heavy mirrors are equivalent to those of a cavity 
with low power and light mirrors if the two cavities have the same $\sigma$. 

The parameter $\sigma$ has a simple geometric interpretation. From the 
equilibrium condition, eq.~(\ref{equilx}), we see that the maximum increase 
in the cavity length due to radiation pressure is $\sigma/k$. 
Therefore, $\sigma/\pi$ is the maximum number of fringes the mirrors can be 
pushed apart by the radiation pressure. 

We can think of $\sigma$ as a function of the incident power
\begin{equation}
   \sigma = \frac{2 k G_{cav}}{c \mu \omega_0^2} P_{in},
\end{equation}
where $G_{cav}$ is the cavity gain, eq.~(\ref{Gcav}). 
Therefore, the simplest way to change the 
parameter $\sigma$ is to vary the input power of the Fabry-Perot cavity.

\subsection{Equilibrium Condition}

The equilibrium states depend on properties of Airy function:
\begin{equation}
   A(\phi) = \frac{1}{1 + F \sin^2 \phi}.
\end{equation}
A discussion of Airy function can be found in ref.~\cite{Born:1980}. 
Some of the 
properties of Airy function are given in the Appendix~\ref{Airy}. 

Let us introduce the phase variables for the cavity length and the control 
parameter: 
\begin{eqnarray}
  \phi & = & k z , \\
  \eta & = & k u .
\end{eqnarray}
In terms of the phase 
variables the condition for equilibrium, eq.~(\ref{equilx}), becomes 
\begin{equation}\label{equilCond}
   \frac{1}{\sigma} (\phi - \eta) = A(\phi).
\end{equation}
By solving this equation we can find  equilibrium states, $\phi$, for any  
given value of the control parameter, $\eta$. 
Even more important is an inverse application of the equilibrium 
condition. Namely, we can turn any state, $\phi$, into an equilibrium 
state by adjusting the control parameter, $\eta$, according to 
\begin{equation}\label{invAppl}
   \eta = \phi - \sigma A(\phi) .
\end{equation}

An operation point for the Fabry-Perot cavity is an equilibrium state, at 
which we want to maintain the cavity. We select the operation point by 
specifying the phase, $\phi_0$, and by adjusting the control 
parameter according to the eq.~(\ref{invAppl}). 

In LIGO interferometers the operation point is chosen at the maximum of Airy 
function $(\phi_0 = \pi n)$, where $n$ is integer. To obtain an equilibrium 
state at these points we must have the control parameter set to the following 
values 
\begin{equation}\label{maxP}
   u = \frac{\lambda}{2} \left( n - \frac{\sigma}{\pi} \right) .
\end{equation}
Since all the peaks of Airy function are equivalent the different values 
of the control parameter, which correspond to different values of $n$, 
are all equivalent. The least absolute value of the control parameter 
corresponds to $n = \mathrm{round}(\sigma/\pi)$. 
Since the control parameter is a linear combination of the actuator force 
and the displacement of the suspension point, eq.(\ref{controlPar}),  
this condition can be achieved in two ways. 
We can either apply the forces to the mirrors or shift the suspension points.
The larger the actuator forces the higher the noise in the actuators. 
Therefore, it is desirable to operate the cavities with as less force on 
the mirrors as possible.

\subsection{Graphical Construction of Equilibrium States}

The equilibrium states are given by the nonlinear equation, 
eq.~(\ref{equilCond}), which cannot be solved analytically. 
However, we can analyze the equilibrium states using a graphical 
construction. 
The graphical method for finding the equilibrium states appears in 
several papers, for example, in ref.~\cite{Marburger:1978}. 
In our approach we use the graphical method not only to show approximate 
locations of the equilibrium states but also to derive exact statements 
about equilibrium and stability of the Fabry-Perot cavity. 

\begin{figure}[ht]
\begin{center}
  \caption{Graphical Construction of Equilibrium States}
  \label{graphRep}
  \includegraphics[width=8cm]{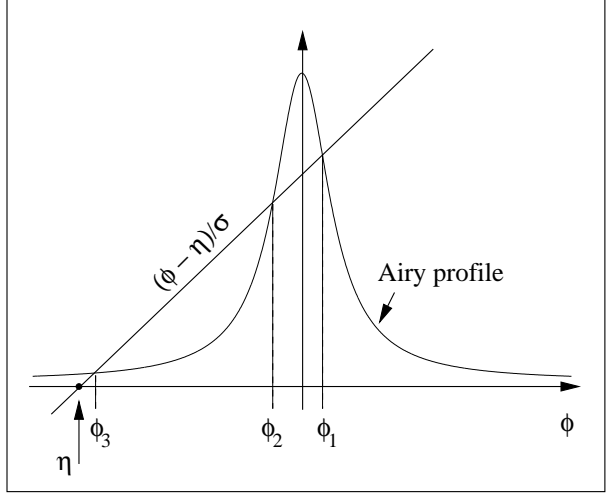}
\end{center}
\end{figure}

The graphical construction is to plot the functions in 
the left and the right sides of the equation~(\ref{equilCond}) and look for 
their intersections. 
The left side of the equation is the pendulum restoring force in 
dimensionless units. It corresponds to a straight line shown in 
Fig.~\ref{graphRep}. The line has the slope $\sigma^{-1}$ and the 
intercept $\eta$. The right side of the equation 
is the Airy function, which is the radiation pressure normalized to unity 
at its maximum. 
The points, where the straight line intersects the Airy function,  
are the equilibrium states. In Figure~\ref{graphRep} there are 
three equilibrium states: $\phi_1, \phi_2$ and $\phi_3$.

\subsection{Condition for Multistability}

A Fabry-Perot cavity is multistable if it has more than one equilibrium state, 
which belong to the same value of the control parameter. 
To derive a condition for multistability consider a line 
tangential to the Airy profile at some point, $\phi_0$. 
The slope of the tangent line is $A'(\phi_0)$, the 
derivative of Airy function at this point. 
The highest slope of the tangent line corresponds to the maximum value of 
the derivative 
\begin{equation}\label{maxSlope}
   \mathrm{max} \{ A' \} \approx \frac{3}{8}\sqrt{3F},
\end{equation}
which is derived in Appendix~\ref{Airy}.
If the slope of the straight line, which represents the pendulum 
restoring force, is greater than the maximum slope, 
the line intersects the Airy profile at only one point. 
Otherwise, the straight line 
and Airy profile can have several common points. 
Therefore, there is a critical value of the parameter $\sigma$, for which 
a qualitative change in the equilibrium states occurs. 
This critical value is defined by the maximum slope 
\begin{equation}\label{sigmaCr}
   \sigma_{cr} \equiv \frac{1}{\mathrm{max} \{ A' \}} 
   \approx \frac{8}{3\sqrt{3F}},
\end{equation}
and depends only on the finesse of the Fabry-Perot cavity. 
These geometric arguments lead us to the condition for multistability. 
The Fabry-Perot cavity is 
\begin{center}
\begin{tabular}{ll}
   stable, if       & $\sigma < \sigma_{cr}$ \\
   multistable, if  & $\sigma > \sigma_{cr}$.
\end{tabular}
\end{center}

LIGO Fabry-Perot cavities are multistable: $\sigma = 3.9$, which is 
greater than the critical value, $\sigma_{cr} = 0.012$.

\subsection{Higher order bifurcations}

\begin{figure}[ht]
\begin{center}
   \caption{Pitchfork bifurcation}
   \label{pitchfork}
   \includegraphics[width=8cm]{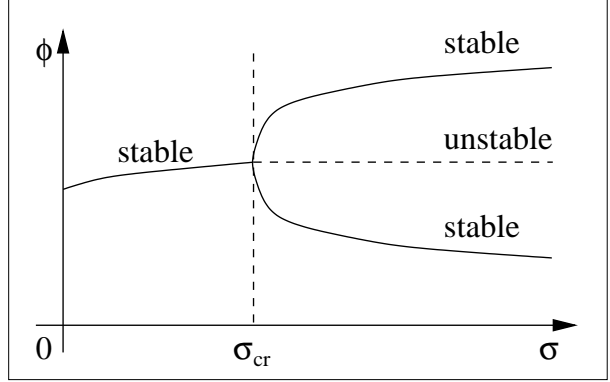}
\end{center}
\end{figure}

At low radiation pressure ($\sigma < \sigma_{cr}$) the cavity has only 
one equilibrium state. If we increase the radiation pressure and reach 
the critical point ($\sigma = \sigma_{cr}$) the 
equilibrium state splits into three states, as shown in Fig.~\ref{pitchfork}. 
Two of the new states 
are stable and one is unstable. Such a phenomenon is common in nonlinear 
physics and is called the pitchfork bifurcation \cite{Rowlands:1990}, 
\cite{Andronov:1966}. 

\begin{figure}[ht]
\begin{center}
   \caption{Higher order bifurcations}
   \label{bifurc}
   \includegraphics[width=8cm]{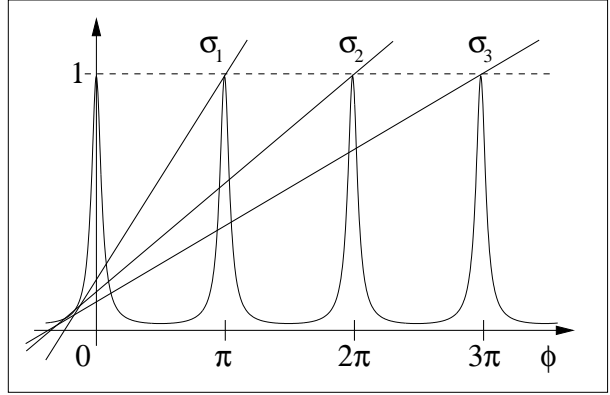}
\end{center}
\end{figure}

If we keep increasing the radiation pressure, at some point 
another bifurcation 
occurs and more equilibrium states are created. We can continue the 
process and obtain 
an infinite number of bifurcations in the Fabry-Perot cavity. 
The values of the parameter $\sigma$, at which these bifurcations occur,  
can be found from the graphical construction shown in Fig.~\ref{bifurc}. 
These values are $\sigma_n \approx \pi n$ 
where $n = 1,2, \ldots $ is the order of the bifurcation. In terms of 
power the condition for the $n$th order bifurcation to occur is 
\begin{equation}
   P_{max} = \frac{1}{4} c \mu \omega_0^2 \lambda n . 
\end{equation}

Exact number of equilibrium states depends not only on the parameter 
$\sigma$, but also on the control parameter, $u$.

\newpage
\section{Stability and Hysteresis}

\subsection{Multistability of Fabry-Perot cavity}

Fabry-Perot cavity with suspended mirrors is a multistable dynamical 
system. There are multiple equilibrium states in the cavity 
which correspond to the same value of the control parameter. 
Some of them are stable and some are unstable. 
A multistable cavity being brought near unstable equilibrium state 
moves to a nearest stable equilibrium state, which belongs to the 
same value of the control parameter. 

\begin{figure}[ht]
\begin{center}
   \caption{Multistability of cavity length}
   \label{mstabX}
   \includegraphics[width=8cm]{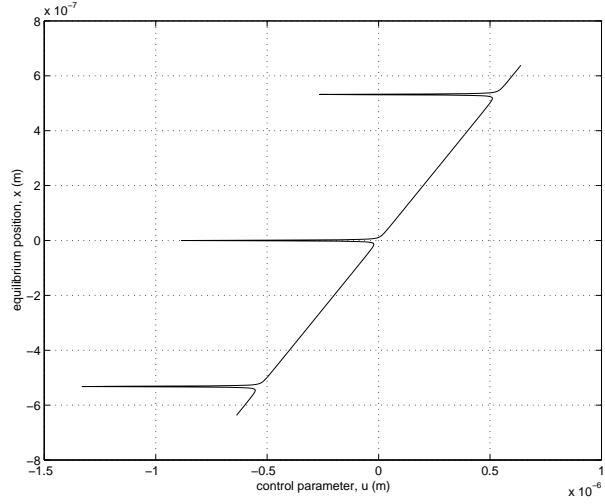}
\end{center}
\end{figure}

Multistability of the Fabry-Perot cavity manifests itself in 
multi-valued functions describing the cavity state. 
For example, the equilibrium cavity 
length, $z$, is a multi-valued function of the control parameter, $u$. 
This function is defined implicitly:
\begin{equation}
   z = u + \frac{\sigma}{k} A(kz), 
\end{equation}
and is shown on Fig.~\ref{mstabX}. 

\begin{figure}[ht]
\begin{center}
  \caption{Multistability of power in Fabry-Perot cavity. 
  The parameters are chosen so that the curve goes through the 
  LIGO operation point}
  \label{mstabP}
  \includegraphics[height=6cm]{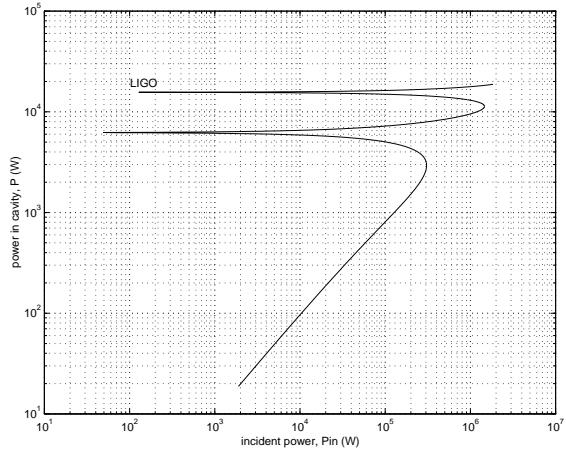}
\end{center}
\end{figure}

A similar multistable response appears if we vary the input power, $P_{in}$, 
within a wide range, but keep the control parameter fixed. This way we 
obtain the power in the cavity, $P$, as a multi-valued 
function of the input power. Such function also is defined implicitly:
\begin{equation}\label{Pmult}
   P = \frac{G_{cav} P_{in}}{1 + F \sin^2  
       \left( ku + \frac{2 k P}{c \mu \omega_0^2} \right)} .
\end{equation}
The plot of this function  for a particular value of the control parameter 
($u = -\frac{2 P_{max}}{c \mu \omega_0^2}$) is shown in Fig.~\ref{mstabP}. 
This is a typical multistability curve similar to those, which 
frequently appear in studies of the bistable optical devices 
\cite{Gibbs:1985}, \cite{Wherrett:1985}.

Both plots above have retrograde segments which correspond to unstable 
equilibrium states of the cavity. In general any state on a 
retrograde part of a multistable curve is unstable \cite{Andronov:1966}. 

The formula which describes the power inside the cavity as 
a function of the input power, eq.~(\ref{Pmult}), is equivalent to the 
corresponding formula for the fixed length cavity filled 
with a non-linear medium. The index of refraction of such medium has 
a quadratic nonlinearity 
\begin{equation}
    n(E) = n_0 + n_2 |E|^2,  
\end{equation}
where $n_0$ and $n_2$ are the coefficients of the nonlinear index of 
refraction. A review of the extensive research which had been done on the
cavities with the quadratic nonlinearity can be found in \cite{Gibbs:1985}.
Here we obtain an exact condition for this equivalence.
In the equation~(\ref{Pmult}) the phase is a function of the power:
\begin{equation}
   \phi = k u + \frac{2 k}{c \mu \omega_0^2} |E|^2.
\end{equation}
A similar dependence 
appears in the cavity with fixed length but filled with the nonlinear 
medium. The equivalence is achieved by tuning the parameters according to
the equation:
\begin{equation}\label{n2}
    n_2 l_0 = \frac{2 k}{c \mu \omega_0^2},
\end{equation}
where $l_0$ is the length of the cavity with the nonlinear medium.
Therefore, equilibrium states of a cavity with suspended mirrors are 
equivalent to those of a cavity filled with the nonlinear medium 
if the condition, eq.~(\ref{n2}) is satisfied. This equivalence 
allows us to extend some results obtained for the cavities with the medium 
to the cavity with suspended mirrors.

\subsection{Hysteresis of Mirror Position}

The instability causes jumps and a hysteresis to appear. In this section we 
describe the mirror jumps and the hysteresis in the 
mirror position in the Fabry-Perot cavity. 

\begin{figure}[ht]
\begin{center}
   \caption{Hysteresis and instability region}
   \label{hyst}
   \includegraphics[height=6cm]{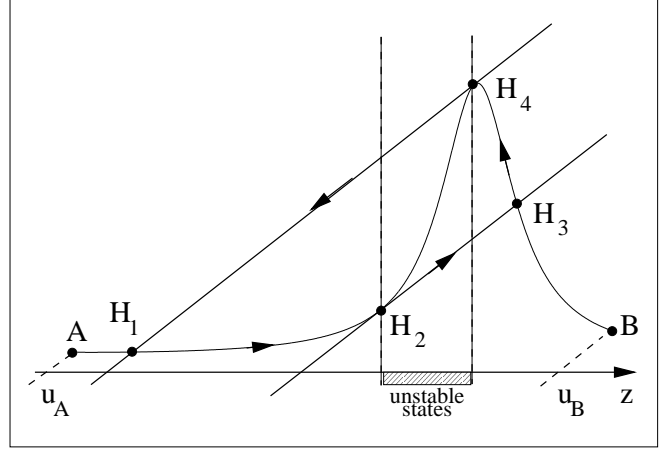}
\end{center}
\end{figure}

Assume that one mirror is at rest and the other is moving. Let this motion 
be so slow that the cavity remains in equilibrium. 
We can produce such motion by a slow change of the control parameter.
The motion of the mirror is shown on the parameter-state diagram, 
Fig.~\ref{hyst}.

Let ``$A$'' be a starting point. Assume that we force the mirror to move
by changing the control parameter from $u_A$ to $u_B$.
As we increase the value of the control 
parameter the mirror moves steadily from point ``$A$'' through point $H_1$ 
to point $H_2$. As the mirror reaches point $H_2$ a sudden transition to 
point $H_3$ occurs. This jump happens because both points $H_2$ and $H_3$ 
belong to the same value of the control parameter but point $H_2$ is unstable. 
After the mirror reaches point $H_3$, it resumes its steady motion and reaches 
the point ``$B$''. The sequence of states is 
\begin{equation}
   A \rightarrow H_1 \rightarrow H_2 \rightarrow H_3 \rightarrow B .
\end{equation}
The jump $H_2 \rightarrow H_3$ is accompanied by a rapid build-up 
of power in the cavity, and, therefore, takes time of the order of the 
storage time to complete. 

In the reverse process, we change the control parameter from $u_B$ to 
$u_A$ and thereby force the mirror to move in the opposite direction.
On the way back the mirror does not retract its path. 
Instead, it follows a sequence 
\begin{equation}
   B \rightarrow H_3 \rightarrow H_4 \rightarrow H_1 \rightarrow A .
\end{equation}
The reverse motion of the mirror also has a jump but at a different place: 
point $H_4$. The transition $H_4 \rightarrow H_1$ is accompanied by 
a rapid drop of power in the cavity. 
Thus a slow sweep of the mirror across the resonance is an irreversible 
process, known as hysteresis. The complete cycle:
\begin{equation}
   H_1 \rightarrow H_2 \rightarrow H_3 \rightarrow H_4 \rightarrow H_1 ,
\end{equation}
is the hysteresis loop, which is typical for nonlinear systems.

A similar hysteresis appears if we keep the control parameter fixed but 
vary the input power. Such hysteresis was observed by Dorsel and 
collaborators and is described in ref.\cite{Dorsel:1983}.

\subsection{Boundaries of Instability}

The states on the retrograde segments of the multistability curve on
Fig.~\ref{mstabX} and Fig.~\ref{mstabP} are unstable. These states form an
instability region which is confined by the two tangent lines with the slope
$\sigma^{-1}$, as shown in Fig.~\ref{hyst}. The boundaries of the
instability region are defined by the equation
\begin{equation}\label{boundStab}
   A'(\phi) = \frac{1}{\sigma}.
\end{equation}
There are two solutions of this equation, which are  
left and right boundaries of the instability region. 
In the large-$F$ approximation the instability region is 
\begin{equation}\label{instReg}
   - \left( \frac{2 \sigma}{F} \right)^{\frac{1}{3}} < \phi_0 < 
   - \frac{1}{2 \sigma F} . 
\end{equation}

The left boundary corresponds to low power inside the 
Fabry-Perot cavity. The right boundary is located very close 
to the peak of the Airy function and corresponds to high power. 
For example, in the 4km LIGO interferometers the right boundary 
of the instability is $\phi_0 = -7.5 \mu$rad. 
Thus the separation between the right boundary 
and the peak is very small, roughly a thousandth of the width 
of Airy function.

\newpage
\section{Hamiltonian Dynamics}

\subsection{Conditions for Adiabatic Motion}

Motion of the mirrors belongs to the adiabatic regime only if it is slow;
the characteristic time of such motion must be much greater than the cavity
storage time. We define the characteristic time differently for small- and
large-amplitude mirror motions.

We consider the amplitude small if it is less than the width of 
the resonance. Usually such motion is oscillatory. 
Then the characteristic time is equal to the period of these 
oscillations, $T_{\mathrm{osc}}$. Therefore, the motion is adiabatic if 
$T_{\mathrm{osc}} \gg \tau$.

Large-amplitude mirror motion takes place when the changes in the cavity 
length are greater than the width of resonance, $x_w$. During 
such motion the relative velocity of the mirrors, $v$, can be considered 
constant within the time the mirrors pass through a resonance. Then 
the characteristic time is the time it takes for the length to change by the 
width of one resonance: $x_w/v$. 
Therefore, the motion is adiabatic if $x_w/v \gg \tau$.
This condition can also be seen as a requirement on the relative 
velocity of the mirrors. Namely, the large-amplitude mirror 
motion is adiabatic if $v \ll v_{ad}$, where $v_{ad}$ 
is the adiabatic threshold velocity:  
\begin{equation}
   v_{ad} \equiv \frac{x_w}{\tau} \approx \frac{c \lambda}{\pi F L_0} .
\end{equation}
The threshold velocity depends only on the parameters of the cavity 
and the wavelength of the laser. For LIGO Fabry-Perot cavities 
$v_{ad} = 7.4 \times 10^{-7}$ m/s.

\subsection{Effective Potential}

In the adiabatic regime the field inside the cavity, $E(t)$, is 
completely defined by the length, $z(t)$, see eq.~(\ref{Eadiab}). 
Therefore, we can eliminate the field 
and consider only the dynamics of the length. The equation for 
the length dynamics becomes 
\begin{equation}\label{xadiab}
   \ddot{z} + \gamma \dot{z} = - \omega_0^2 (z - u) + \frac{1}{\mu}
      R_{\mathrm{max}} A(kz),
\end{equation}
where $\mu$ is the reduced mass and $R_{\mathrm{max}}$ 
is the maximum radiation pressure defined by eq.~(\ref{Rmax}).
This equation describes the relative motion of the mirrors driven by  
the pendulum restoring force and the radiation pressure. 
A similar equation for the Fabry-Perot cavity with one suspended mirror
appeared earlier in refs.~\cite{Dorsel:1983}, 
\cite{Deruelle:1984}, \cite{Tourrenc:1985}. 

The pendulum restoring force is generated by the usual quadratic potential 
\begin{equation}\label{Vpend}
   V_p(x) = \frac{1}{2} \mu \omega_0^2 (z - u)^2 ,
\end{equation}
which depends on the control parameter, $u$. 
The force due to the radiation pressure is generated by a potential
\begin{equation}\label{Vrad}
   V_{\mathrm{rad}}(z) = - R_{\mathrm{max}}
        \frac{\arctan[ \sqrt{F+1} \tan (k z) ]}{k \sqrt{F+1}}, 
\end{equation}
which is derived in the Appendix~\ref{Airy}. 

The sum of the two forces is generated by the effective potential
\begin{equation}
   V_{\mathrm{eff}}(z) = V_p(z) + V_{\mathrm{rad}}(z) .
\end{equation}
Thus the relative motion of the mirrors  
interacting with the electro-magnetic field in the cavity is equivalent 
to the motion of a particle with the reduced mass in the field of the 
effective potential. 
Such representation is valid only in the adiabatic regime, when the 
delay time can be neglected.
Beyond the adiabatic regime the delay time becomes important and the 
field depends on history of mirror's motion within the storage time. 

A remarkable property of the adiabatic regime is that 
the dynamics is hamiltonian. In other words, we can define a total 
energy for the motion in the effective potential:
\begin{equation}\label{energy}
   {\cal E} = \frac{\mu \dot{z}^2}{2} + V_{\mathrm{eff}}(z).
\end{equation}
Thus the Fabry-Perot cavity in the adiabatic regime is a hamiltonian 
system. Since there is friction ($\gamma > 0$) the energy is not conserved:
\begin{equation}
   \dot{\cal E} = - \gamma \mu \dot{z}^2 < 0.
\end{equation}
However, the friction in suspension systems of the gravitational wave 
detectors is very small. Consequently the 
energy changes very little (less than a part per million) over a period 
of the pendulum. Therefore, on time scales of the order of 
few seconds and even 
minutes we can think that the energy is conserved. 

Note that if there is no friction ($\gamma = 0$) the energy of the 
motion in the effective potential, eq.~(\ref{energy}), 
is conserved but the combined energy 
of the mirrors and the electro-magnetic fields in the cavity may not be 
conserved. The cavity, being multistable, may release or absorb 
energy as it moves from an unstable to a stable state.

\subsection{Global Properties of Effective Potential}

The effective potential as a function of cavity length depends essentially 
on three parameters: $F$, $\sigma$ and $\eta$. This can be seen if 
we write the effective potential in terms of dimensionless parameters:
\begin{equation}
   V_{\mathrm{eff}} = \frac{\mu \omega_0^2}{k^2} 
      \left[ \frac{(\phi - \eta)^2}{2} -
      \sigma \; \frac{\arctan( \sqrt{F+1} \tan \phi)}{\sqrt{F+1}} \right].   
\end{equation}
The overall constant
provides correct dimensions to the potential, but otherwise is irrelevant. 
The parameter $\sigma$ and the control parameter $\eta$ can be dynamically 
adjusted, but the coefficient of finesse, $F$, cannot be.  
Therefore, we only consider how 
the effective potential changes with respect to arbitrary changes in 
the two parameters: $\sigma$ and $\eta$. 

\begin{figure}[ht]
\begin{center}
   \caption{Effective potential as function of $\sigma$}
   \label{pShapes}
   \includegraphics[height=6cm]{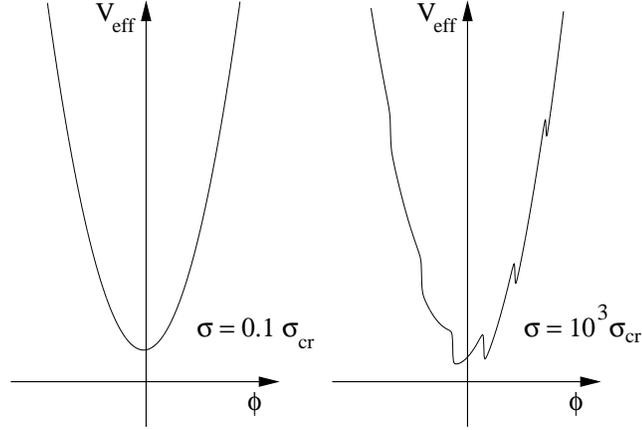}
\end{center}
\end{figure}

Global properties of the effective potential are defined by one 
parameter only, $\sigma$. For $\sigma < \sigma_{cr}$ the potential has only 
one minimum and for $\sigma > \sigma_{cr}$ there are several 
local minima. Effective potential for two different values of $\sigma$ 
is shown in Fig.~\ref{pShapes}. 
The plots correspond to stable and multistable Fabry-Perot cavities. 

The effective potential as a function of two parameters establishes a 
connection between the dynamics of the 
Fabry-Perot cavity in the adiabatic regime and the dynamical systems of 
the catastrophe theory \cite{Arnold:1984}. 
The catastrophe theory studies singularities (abrupt changes) 
in equilibrium states of a 
dynamical system caused by a continuous change of its parameters. 
The pitchfork bifurcation, as described above, is one of the examples of 
such singularities.

\subsection{Minima and Maxima of the Effective Potential}

\begin{figure}[ht]
\begin{center}
   \caption{Minimum and maximum of effective potential}
   \label{minmax}
   \includegraphics[height=8cm]{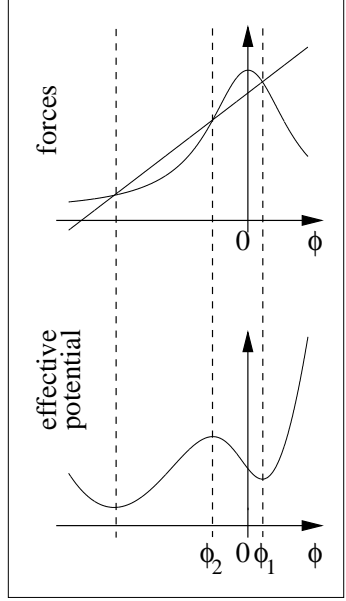}
\end{center}
\end{figure}

The control parameter, $\eta$, allows us to fine tune the effective potential. 
By changing the control parameter we can shift the positions of local 
minima and maxima of the potential. 
To illustrate this consider the following example. 
Let the operation point be $\phi_1$ as shown in Figure~\ref{minmax}.

To make this state an equilibrium we have to adjust the
control parameter according the eq.~(\ref{invAppl}). This is equivalent 
to solving the following equation:
\begin{equation}
   \left. \frac{d V_{\mathrm{eff}}}{d \phi} \right|_{\phi = \phi_1} = 0.   
\end{equation}
Since at the point, $\phi_1$, the effective potential has a local minimum: 
\begin{equation}
   \left. \frac{d^2 V_{\mathrm{eff}}}{d \phi^2} \right|_{\phi = \phi_1} > 0    
\end{equation}
this state is stable.
From the figure we see that if we choose the operation point 
at $\phi_2$, we obtain the same control parameter and arrive 
at the same effective potential. 
However, at the point $\phi_2$ the effective potential has a local maximum:
\begin{equation}
   \left. \frac{d^2 V_{\mathrm{eff}}}{d \phi^2} \right|_{\phi = \phi_2} < 0.    
\end{equation}
Therefore, the equilibrium state at  $\phi_2$, is unstable. 

Thus, stable equilibrium positions correspond to local 
minima and unstable equilibrium positions correspond 
to local maxima of the effective potential. By adjusting the 
control parameter we can turn any state into a stable equilibrium 
if that state is outside the instability region, eq.(\ref{instReg}).

\newpage
\section{Resonance of Cavity Length}

\subsection{Resonant Frequency}

A stable equilibrium state of the cavity corresponds to 
a local minimum of the effective potential. 
Let the minimum of the effective potential be at $z_0$. 
The small-amplitude mirror motion near this equilibrium state 
produces harmonic oscillations of the cavity length. 
The frequency of such oscillations, $\Omega_0$, is defined 
by the curvature of the potential at the minimum: 
\begin{equation}
   \Omega_0^2 = \left. \mu^{-1} \frac{d^2 V_{\mathrm{eff}}}{d z^2} 
      \right|_{z_0},
\end{equation}
The oscillations of length due to motion of a single mirror, 
confined in one of the minima of the effective potential 
were observed by Dorsel et al \cite{Dorsel:1983, Dorsel:1984}. 

\begin{figure}[ht]
\begin{center}
   \caption{Frequency of length oscillations as function of operation point}
   \label{smallosc}
   \includegraphics[width=8cm]{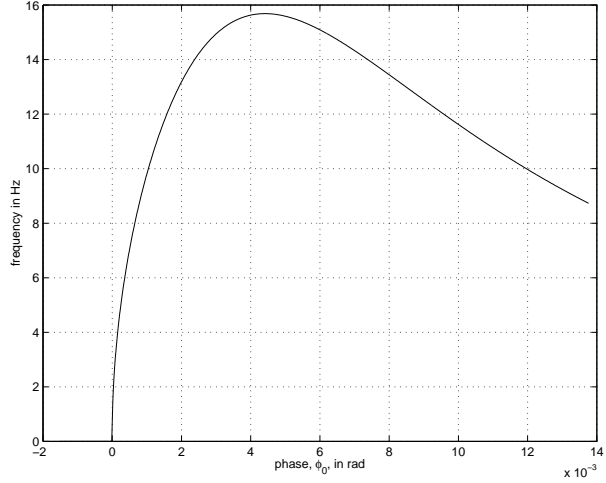}
\end{center}
\end{figure}

Since there are multiple equilibrium states, there are multiple frequencies 
for the oscillations of the cavity length. 
Each minimum of the effective potential has its own frequency. 
These are the resonant frequencies 
of the cavity length, if the cavity is driven by the external sinusoidal 
force. At high radiation pressure these frequencies substantially deviate from 
the pendulum frequency, $\omega_0$.
We find that the frequencies can be expressed entirely in terms of the 
parameter $\sigma$ and the location of the minimum:
\begin{equation}\label{OmegaSqr}
   \Omega_0^2(\phi_0) = \omega_0^2 \; [ 1 - \sigma A'(\phi_0) ],  
\end{equation}
where $\phi_0 = k z_0$ and $A'(\phi_0)$ is the derivative of the 
Airy function (Appendix~\ref{Airy}). 
The plot of $\Omega_0/2\pi$ as a function of the phase $\phi_0$ with 
LIGO parameters is shown in Fig.~\ref{smallosc}.

The frequency reaches maximum at the 
right inflection point of the Airy profile. The maximum frequency is 
\begin{equation}
   \Omega_0 (\phi_{\mathrm{infl}}) = 
      \omega_0 \left( 1 + \frac{\sigma}{\sigma_{cr}} \right)^{\frac{1}{2}} . 
\end{equation}
In LIGO Fabry-Perot cavity the maximum frequency is 15.7 Hz.

\subsection{Adjustment of Resonance Frequencies}

The resonances of cavity length make the Fabry-Perot 
cavity a narrow-band transducer between mechanical and optical signals. 
This observation was made by Dorsel and collaborators 
\cite{Dorsel:1985}. They also suggested that a resonant   
frequency can be tuned by changing the power of the incident laser.
The adjustment of the incident power, which are equivalent 
to the adjustments of the parameter $\sigma$ is not the only way 
to tune the resonant frequency. We can also vary the control parameter. 
By changing the control parameter we can shift the minimum of the effective 
potential and thus change the resonant frequency corresponding to that 
minimum. 

On the other hand we may require that the operation point remains 
fixed as we tune the resonant frequency. This can be done if we vary the two 
parameters, $\sigma$ and $\eta$, simultaneously so that the equilibrium 
condition is preserved. For this we need to accompany any change 
$\delta\sigma$ by the change in the control parameter as follows 
\begin{equation}
   \delta \eta = - A(\phi_0) \; \delta \sigma .
\end{equation}
This way we change the curvature of the effective potential 
at the minimum without changing the location of that minimum. 

\begin{figure}[ht]
\begin{center}
   \caption{Fine tuning of effective potential}
   \label{freqAdj}
   \includegraphics[width=10cm]{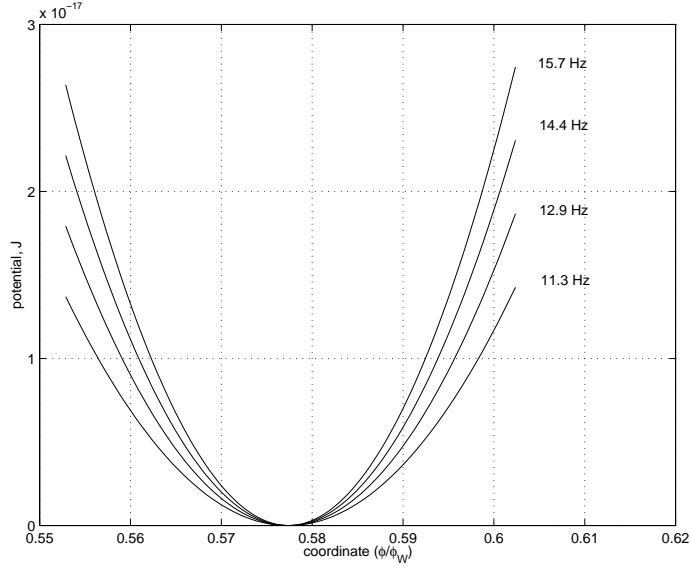}
\end{center}
\end{figure}
The result is the resonant frequency as a function of the control 
parameter 
\begin{equation}
   \Omega_0^2(\eta) = \omega_0^2 \; \left[ 1 - (\phi_0 - \eta) 
   \frac{A'(\phi_0)}{A(\phi_0)} \right]. 
\end{equation}
The fine tuning of the effective potential and the corresponding frequencies 
are shown in Fig.~\ref{freqAdj}.

\subsection{Plateaus of the Effective Potential}

By tuning the parameters, $\sigma$ and $\eta$, we can make the 
effective potential more and more flat at one of its minima and even 
form a plateau: 
\begin{equation}
   \frac{d^2 V_{\mathrm{eff}}}{d z^2} = 0 .
\end{equation}

One way of creating the plateau is to tune the control parameter, $\eta$. 
In this method we flatten the potential by bringing the 
minimum closer to the nearest local maximum. This method of obtaining the 
plateau by merging the local minimum with the nearest local maximum has 
a drawback. Namely, in this method we bring the operation point 
closer and closer to the instability region, eq.~(\ref{instReg}). 

Another way of forming the plateau at the minimum is to start 
with lower power ($\sigma < \sigma_{cr}$) and gradually increase the power 
until the bifurcation occurs ($\sigma = \sigma_{cr}$). 
In this method we begin 
with a stable cavity: the effective potential has only one minimum.  
We have to adjust the control parameter so that the operation point 
coincides with this minimum: 
\begin{equation}
   \frac{d^3 V_{\mathrm{eff}}}{d z^3} = 0 .
\end{equation}
By increasing the incident power 
$\sigma \rightarrow \sigma_{cr}$, we make the potential more and 
more flat at the minimum. At the bifurcation point 
($\sigma = \sigma_{cr}$) the plateau is formed.

\subsection{Period of Anharmonic Oscillations}

The oscillations of length in almost flat potential 
are anharmonic and can have very large period. 
At the critical point the potential is of fourth order: 
\begin{equation}
   V_{\mathrm{eff}}(z) = V_{\mathrm{eff}}(z_0) + 
                 \frac{1}{4!} V_{\mathrm{eff}}^{(4)}(z_0) (z - z_0)^4 .
\end{equation}
The period of the anharmonic oscillations in the fourth order potential is 
\begin{equation}\label{periodAnh}
   T_{\mathrm{osc}} = \Gamma^2\left( \frac{1}{4} \right)\left[\frac{6 \mu}
      {\pi V_{\mathrm{eff}}^{(4)}(z_0)} \right]^{\frac{1}{2}}\frac{1}{a}, 
\end{equation}
where $a$ is the amplitude of the oscillation. Since the amplitude of 
mirror motion and correspondingly the amplitude of the length oscillations is 
very small (of the order of a few $\mu$m), the period can be very large.

In practice it is unlikely that the cavity is maintained exactly at the 
bifurcation point. More likely the long period oscillations are 
obtained  slightly below the critical point 
$\sigma < \sigma_{cr}$. In this case the potential is not quartic, 
nonetheless the oscillations have long period, which can be found from 
a general expression 
\begin{equation}\label{periodGen}
   T_{\mathrm{osc}} = \sqrt{2 \mu} \int\limits_{z_1}^{z_2} 
      \frac{dz}{[ {\cal E} - V_{\mathrm{eff}}(z) ]^{\frac{1}{2}}},
\end{equation}
where $z_1$ and $z_2$ are the turning points of the oscillations. 
Both equations~(\ref{periodAnh}) and (\ref{periodGen}), appear in 
books on Mechanics, for example in ref.~\cite{Landau:Mech}. 

In Table~\ref{lowFreq} we show numerical values of 
frequencies of the oscillations in the nearly flat potential for 
LIGO Fabry-Perot cavities. The flatness of the potential 
is characterized by the deviation of the parameter $\sigma$ from its
critical value. 
\begin{table}[ht]
   \caption{Period of anharmonic oscillations in almost flat potential}
   \label{lowFreq}
   \begin{center}
   \begin{tabular}{cccc}
   \hline
   \hline
    amplitude (m) & \multicolumn{3}{c}{period (s)} 
   \\
   $a$ & $\sigma = 0.96 \; \sigma_{cr}$  
       & $\sigma = 0.98 \; \sigma_{cr}$  
       & $\sigma = \sigma_{cr}$  
   \\ \hline 
   $10^{-8}$  & 1.34 & 1.34 & 1.33 
   \\ 
   $10^{-9}$  & 2.31 & 2.36 & 2.41 
   \\
   $10^{-10}$ & 6.39 & 8.74 & 23.2 
   \\ 
   $10^{-11}$ & 6.68 & 9.44 & 234.2  
   \\ 
   \hline   
   \hline
   \end{tabular}
   \end{center}
\end{table}
The table shows that we can achieve period substantially greater 
than the pendulum period ($2\pi/\omega_0 \approx 1.3$ s). This method of 
obtaining very long period by taking advantage of the radiation pressure 
is an interesting alternative to the very long suspension.

\newpage
\section{Implications for Control System}

\subsection{Pound-Drever Locking Servo}

The operation point of the cavity is a stable equilibrium length which 
corresponds to a minimum of the effective potential. 
Due to the ambient seismic motion the actual state of the cavity 
is constantly moving near the operation point. Often the minimum of 
the effective potential is not deep enough to confine the state. In such 
case a negative feedback control system (servo) 
is used to keep the cavity state within the minimum. The dynamics of the
cavity with the control system is described by the same equations as 
before but the control parameter, $u$, becomes a function of the 
cavity length.
In LIGO interferometers the control function, $u(z)$, is proportional to 
the Pound-Drever signal, eq.~(\ref{VPD}):
\begin{equation}\label{signalPD}
   u(z) \propto \mathrm{Im} \{ E(t) \} .  
\end{equation}
Such control function depends on the cavity length $z$ and does not depend 
on the cavity center of mass. 

There is no explicit analytic expression for the function $u(z)$, given by
eq.~(\ref{signalPD}), which is valid for all dynamical regimes.
However, in the adiabatic regime we can obtain a simple formula 
\begin{equation}\label{adiabPD}
   u(z) = - \left( \frac{G}{2k} \right) 
          \frac{\sin 2kz}{1 + F \sin^2 kz} + u_0,
\end{equation}
where $G$ is the gain and $u_0$ is the dc-bias of the control system. 
The minus sign in eq.~(\ref{adiabPD}) accounts for the negative feedback.

If the deviation of the cavity length from the nearest fringe 
is much less than the width of the resonance the control signal becomes 
linear:
\begin{equation}\label{linearPD}
   u(z) \approx - G \left( z - n \frac{\lambda}{2} \right) + u_0 , 
\end{equation}
where $n$ is the order of the fringe. Since the amplitude of motion of 
the mirrors is small (of the order of a micron) the fringe 
order number is a small number: $n = 0, \pm 1, \pm 2, \ldots$.

\subsection{Multistability in Presence of Servo}

Since the control law eq.~(\ref{adiabPD}) is a periodic function of the 
cavity length the cavity with the servo is multistable.
To study multistability in dynamics of the cavity with the 
control system we introduce the static servo potential 
\begin{eqnarray}
   V_{\mathrm{svo}}(z) & \equiv & - \mu \omega_0^2 \int\limits_0^z u(z') dz' \\
      & \propto & \ln (1 + F \sin^2 kz).
\end{eqnarray}
Therefore, in the adiabatic regime the dynamics of the cavity with the 
servo can be described by the combined potential 
\begin{equation}
   V_{\mathrm{eff}} + V_{\mathrm{svo}} = 
      \frac{\mu \omega_0^2}{k^2} \left[ \frac{(\phi - \eta)^2}{2} -
      \sigma \; \frac{\arctan( \sqrt{F+1} \tan \phi)}{\sqrt{F+1}} - 
      \frac{G}{2F} \ln A(\phi) \right] , 
\end{equation}
where the dc-bias is conveniently assigned to the 
pendulum potential ($\eta = k u_0$). 

\begin{figure}[ht]
\begin{center}
  \caption{Combined radiation pressure and servo potential}
  \label{servoPot}
  \includegraphics[height=6cm]{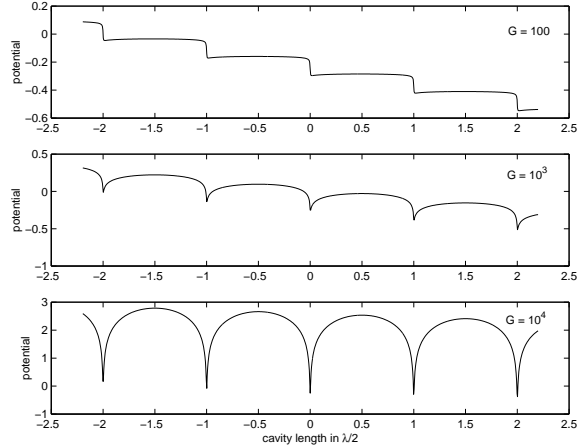}
\end{center}
\end{figure}

For large-amplitude mirror motion the control function, $u(z)$, 
is nonlinear and, therefore, the servo acts as a 
nonlinear device. The effect of the servo depends on 
the gain, $G$, and can be made stronger than the effect of the radiation 
pressure.
Fig.~\ref{servoPot} shows the sum of the radiation pressure potential 
and the servo potential for three different values of the servo gain: 
$G = 10^2, 10^3, 10^4$. 
On the first plot the sum of the two potentials looks like a staircase 
which is characteristic of the radiation pressure potential. Here the 
effect of the servo is negligible compared to the effect of the radiation 
pressure.
On the second plot the ``staircase'' shows dips created by the servo. 
Here the effect of the servo is 
comparable to the effect of radiation pressure. 
On the third plot the dips are bigger than the steps, and 
the servo dominates the radiation pressure.

\subsection{Equilibrium and DC-bias}

Stable equilibrium states of the cavity with the servo 
are the minima of the combined potential:
\begin{equation}
   \frac{d}{d z} (V_{\mathrm{eff}} + V_{\mathrm{svo}}) = 0 . 
\end{equation}
This condition can also be seen as a requirement on the dc-bias. Namely,
a cavity state, $z_0$, becomes an equilibrium if
the dc-bias is set according to the equation: 
\begin{equation}\label{dc-bias}
   u_0 = z_0 - \frac{\sigma}{k}
         \left( 1 - \frac{G}{2\sigma} \sin 2 k z_0 \right) A(kz_0). 
\end{equation}
From this equation we see that, in general, the dc-bias is connected to the 
gain of the servo. Therefore, changes of the servo gain should be 
accompanied by the corresponding changes of the dc-bias. Otherwise the 
equilibrium condition will be lost and the equilibrium state will change.
There is one exception to this rule. In the case when the operation point
is chosen at the peak of the Airy function ($z_0 = \frac{\lambda}{2}$) 
the dc-bias becomes independent of the gain:
\begin{equation}
   u_0 = n \frac{\lambda}{2} - \frac{\sigma}{k} . 
\end{equation}
Note the dependence of the dc-bias on the fringe number, $n$.

\subsection{Stability Provided by Servo}

The negative feedback control system can turn an 
unstable state of the cavity into a stable one. It can also provide greater 
stability margins to the already stable cavity. These effects of the servo 
can be analyzed with the help of the servo potential.

The frequency of the cavity length oscillations in the presence of 
the servo is 
\begin{eqnarray}
   \Omega_0^2 & \equiv & \mu^{-1} \frac{d^2}{d z^2} 
      (V_{\mathrm{eff}} + V_{\mathrm{svo}}) \\
      & = & \omega_0^2 \left\{ 1 - \sigma A'(\phi_0) + 
            G [1 - (F + 2) \sin^2 \phi_0] A^2(\phi_0) \right\} ,  
\end{eqnarray}
By adjusting the gain we can vary this frequency in a wide range. 
We can also turn the unstable state ($\Omega_0^2 < 0$) into a 
stable one ($\Omega_0^2 > 0$).
In particular we can make stable any state of the cavity within a width of 
the fringe ($|\phi_0|< \phi_w$). This can be done by choosing a sufficiently 
high servo gain. For LIGO parameters such gain is $G \approx 1800$.

\newpage
\section{Conclusions}

The suspended mirrors in the Fabry-Perot cavity are described as a pair of
harmonic oscillators driven by nonlinear forces of the radiation pressure.
The oscillators are not independent; they interact with each other through
the laser circulating in the cavity. The normal coordinates for the coupled
oscillators are the position of their center of mass and the cavity length,
both affected by the delay in the cavity. The radiation pressure has almost
no effect on the motion of the center of mass, but affects strongly the
dynamics of the cavity length. To understand universal properties of the
Fabry-Perot cavity as a nonlinear dynamical system we introduced two
parameters. These parameters determine qualitative behavior of the dynamics
of the cavity length common to all Fabry-Perot cavities with suspended
mirrors. The first parameter determines whether the cavity is multistable or
not. There is a critical value for this parameter, at which the bifurcation
occurs, and the cavity becomes multistable. The second parameter is the
generalized control force. It determines the locations of the equilibrium
states. If the control parameter varies within a wide range the equilibrium
state of the cavity follows a hysteresis loop. We analyze stability of the
equilibrium states and identify the instability region. The instability is
explained in terms of the effective potential: stable states correspond to
local minima of the effective potential and unstable states correspond to
local maxima. Each minimum of the effective potential defines the resonance
frequencies of the cavity length oscillations. These resonances make the
cavity a narrow band detector of any disturbances which affect the optical
path of the laser in the cavity, such as mechanical vibrations of the
mirrors. The resonance frequencies can be tuned by changing both of the
parameters.

The nonlinear dynamics, described in this paper, provides grounds for
building a control system for the Fabry-Perot cavities in the interferometric
gravitational detectors. Our results can also be used in the studies of the
Fabry-Perot cavities near the quantum limit.

\appendix
\newpage
\section{Parameters of LIGO Fabry-Perot Cavities}\label{cavPar}

The following table shows the transmissivities and the losses for the 40m and 
4 km LIGO mirrors.  
\begin{table}[ht]
   \caption{Transmissivity and reflectivity of LIGO mirrors}
   \begin{center}
   \begin{tabular}{llll}
   \hline
   \hline
   Mirror \#    & parameter & 40m Prototype & 4km LIGO \\ 
   \hline  
   input mirror & $t_a$ & $0.0758$   & $0.1732$   \\
                & $r_a$ & $0.9971$   & $0.9849$   \\
   \hline  
   end mirror   & $t_b$ & $0.0035$   & $0.0045$   \\
                & $r_b$ & $0.999944$ & $0.999965$ \\
   \hline   
   \hline
   \end{tabular}
   \end{center}
\end{table}

Other parameters for LIGO cavities are listed in the following table 
\begin{table}[ht]
   \caption{Parameters of 40m and 4km LIGO cavities}
   \begin{center}
   \begin{tabular}{llll}
   \hline
   \hline
   Parameter (units)            & Symbol 
                                & 40m Prototype 
                                & 4km LIGO \\ \hline
   laser wavelength (m)         & $\lambda$ 
                                & $5.14 \times 10^{-7}$ 
                                & $1.06 \times 10^{-6}$ \\
   cavity length (m)            & $L_0$  
                                & $38.5$ 
                                & $4000$ \\
   delay time (s)               & $T$ 
                                & $1.28 \times 10^{-7}$ 
                                & $1.33 \times 10^{-5}$ \\
   storage time (s)             & $\tau$ 
                                & $8.60 \times 10^{-5}$ 
                                & $1.74 \times 10^{-3}$ \\ 
   free spectral range (Hz)     & $f_{\mathrm{FSR}}$ 
                                & $3.89 \times 10^{6}$ 
                                & $3.75 \times 10^{4}$ \\ 
   cavity low-pass freq. (Hz)   & $f_{cav}$  
                                & $1.85 \times 10^3$ 
                                & $91.2$ \\ 
   cavity gain                  & $G_{cav}$ 
                                & $645$ 
                                & $130$ \\
   coefficient of finesse       & $F$ 
                                & $4.47 \times 10^{5}$ 
                                & $1.71 \times 10^{4}$ \\
   phase width (rad)            & $\phi_w$ 
                                & $1.49 \times 10^{-3}$ 
                                & $7.64 \times 10^{-3}$ \\
   displacement width (m)       & $x_w$ 
                                & $1.22 \times 10^{-10}$ 
                                & $1.29 \times 10^{-9}$ \\
   Finesse                      & Finesse 
                                & $1051$ 
                                & $205$ \\
   mirror mass (kg)             & $m_a$ 
                                & $1.53$ 
                                & $10.8$ \\
   pendulum frequency (Hz)      & $\omega_0/2\pi$ 
                                & $1.0$ 
                                & $0.77$ \\
   \hline
   \hline
   \end{tabular}
   \end{center}
\end{table}

\newpage
\section{Derivation of Field Equations}\label{defE}

We approximate the laser by a plane electro-magnetic wave with electric
field, $f(x,t)$, which is a solution of the 1-dimensional wave equation:
\begin{equation}
   \frac{1}{c^2} \frac{\partial^2 f}{\partial t^2} - 
   \frac{\partial^2 f}{\partial x^2} = 0.
\end{equation}
Let $f_1(x,t)$ be the electric field corresponding to the right-moving wave
in the cavity and $f_2(x,t)$ be the electric field corresponding to the
left-moving wave. We define the complex amplitudes at the reference planes,
located at $x = 0$ and $x = L_0$. The amplitudes of the right-moving wave,
$E_1$ and $E'_1$, are defined by the following equations:
\begin{eqnarray}
   f_1(0,t)   & = & E_1(t)  e^{i \omega t},\\
   f_1(L_0,t) & = & E'_1(t) e^{i \omega t}.
\end{eqnarray}
Similarly, we define the amplitudes,$E_2$ and $E'_2$, for the left-moving
wave:
\begin{eqnarray}
   f_2(0,t)   & = & E'_2(t) e^{i \omega t},\\
   f_2(L_0,t) & = & E_2(t)  e^{i \omega t}.
\end{eqnarray}
Since the wave functions satisfy the wave equation their values at 
different locations can be obtained by delay:
\begin{eqnarray}
   f_1(L_0,t) & = & f_1(0,  t - L_0/c),\\
   f_2(0,t)   & = & f_2(L_0,t - L_0/c).
\end{eqnarray}
Therefore, the amplitudes at the two reference planes are related to each 
other:
\begin{eqnarray}
   E'_1(t) & = & E_1(t - L_0/c) e^{-i k L_0},\\
   E'_2(t) & = & E_2(t - L_0/c) e^{-i k L_0}.
\end{eqnarray}
If we assume that
\begin{equation}
   L_0 = N \frac{\lambda}{2},
\end{equation}
where $N$ is an integer, we obtain the equations~(\ref{Ep1}-\ref{Ep2}).

The equation for the reflection off the end mirror can be obtained from 
the condition of continuity of the wave function at the mirror surface:
\begin{equation}
   f_2(L_0 + x_b,t) = - r_b f_1(L_0 + x_b,t).
\end{equation}
Therefore, at the reference plane $x = L_0$:
\begin{equation}
   f_2(L_0,t + x_b/c) = - r_b f_1(L_0,t - x_b/c).
\end{equation}
This equation can now be written in terms of the amplitudes:
\begin{equation}
   E_2 (t + x_b/c) = - r_b E'_1(t - x_b/c) e^{-2i k x_b}.
\end{equation}
If we neglect the small time delay due to the mirror motion, $x_b/c$,
we obtain the equation~(\ref{E2}).
Other equations can be derived similarly.

\newpage
\section{Some properties of Airy function}\label{Airy}

\begin{figure}[ht]
\begin{center}
  \caption{Airy Profile}
  \label{airyFs}
  \includegraphics[height=6cm]{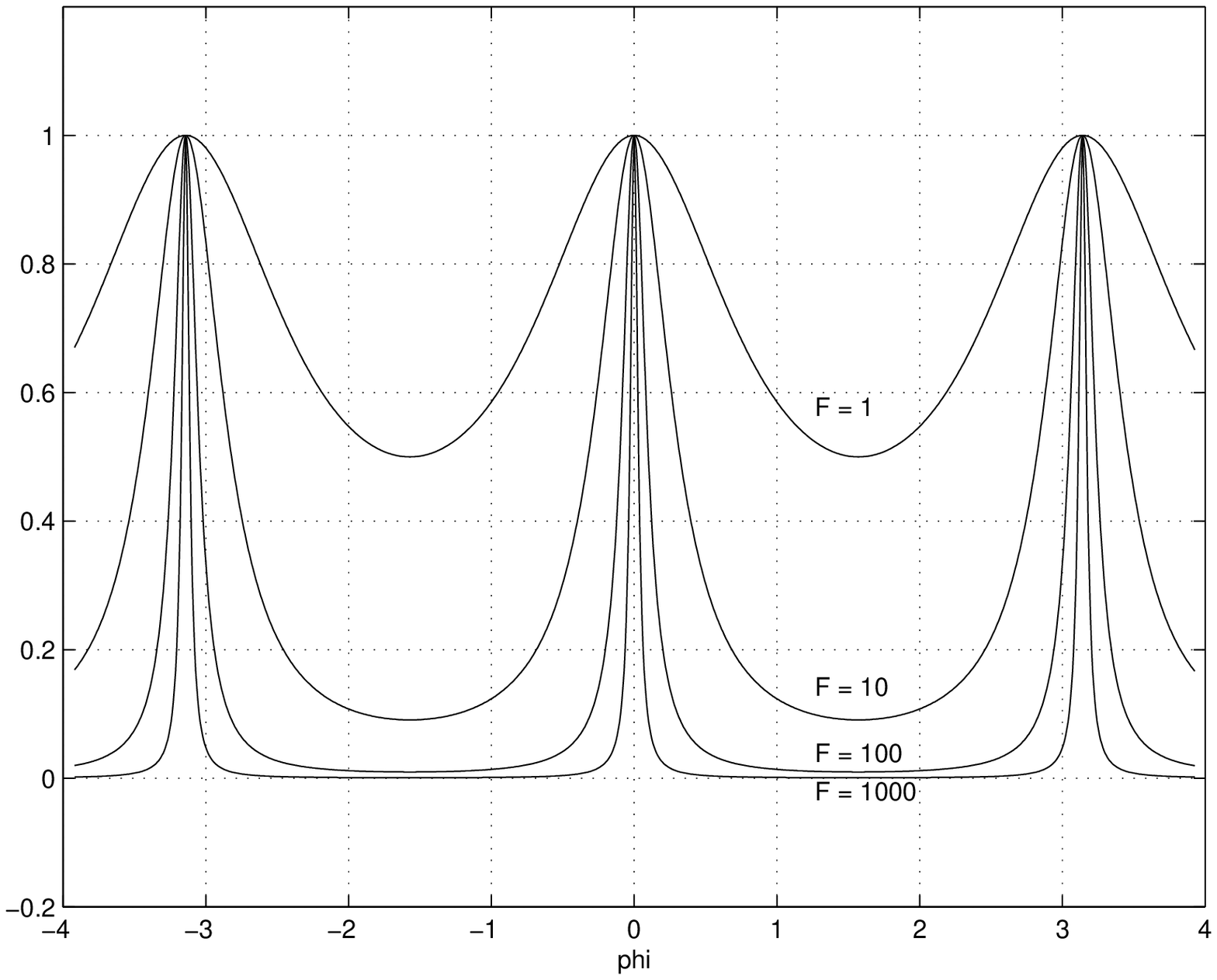}
\end{center}
\end{figure}

Here we present few basic properties Airy function, which are used 
throughout this paper. The Airy function, 
\begin{equation}
   A(\phi) = \frac{1}{1 + F \sin^2 \phi},  
\end{equation}
is a postitive, periodic function of $\phi$ with the period $\pi$. 
The Airy function depends on one parameter, $F$, a positive number called 
the coefficient of finesse. Fig.~\ref{airyFs} shows Airy functions with 
different coefficients of finesse.

The half-width, $\phi_w$, is the phase, at which the Airy function 
equals to half of its maximum value 
\begin{equation}
   \phi_w \equiv \arcsin \frac{1}{\sqrt{F + 2}} \approx \frac{1}{\sqrt{F}}.
\end{equation}
The full width is equal to $2 \phi_w$. Finesse is the ratio of the separation between the peaks to the full width of the peak 
\begin{equation}
   \mathrm{Finesse} \equiv \frac{\pi}{2 \phi_w} 
      \approx \frac{\pi}{2} \sqrt{F}.
\end{equation}
A slope of a tangent line to Airy function at point $\phi$ is defined 
by the derivative of Airy function 
\begin{equation}
   A'(\phi) = - \frac{F \sin 2 \phi}{(1 + F \sin^2 \phi)^2}.   
\end{equation}
The inflection points of Airy function can be found by solving 
the equation: $A''(\phi) = 0$. 
They are located at $\pm \phi_{\mathrm{inf}}$, where 
\begin{equation}\label{Infl}
   \cos 2 \phi_{\mathrm{inf}} = \frac{1}{2F} 
      \left(\sqrt{9F^2 + 4F + 4} - F - 2 \right). 
\end{equation}
The greatest slope the tangent line achieves at the left 
inflection point $\phi = - \phi_{\mathrm{inf}}$. This maximum slope is  
\begin{equation}\label{maxdA}
    A'(- \phi_{\mathrm{inf}}) = \frac{ \left[ \frac{1}{2}(F + 2) 
      \sqrt{9F^2 + 4F + 4} - \frac{3}{2}F^2 - 2F - 2 \right]^{1/2}}{\left( 
      \frac{1}{4}\sqrt{9F^2 + 4F + 4} - \frac{3}{4}F - \frac{3}{2}\right)^2}.
\end{equation}
The expression for the inflection points, eq.~(\ref{Infl}), and the 
maximum slope, eq.~(\ref{maxdA}), are quite complicated, therefore, it 
useful to find simple approximate formulas for these quantities. For 
high finesse cavities (large $F$- limit) such approximations can be 
easily found: 
\begin{eqnarray}
   \phi_{\mathrm{inf}} & \approx & \frac{1}{\sqrt{3 F}},\\
   A'(- \phi_{\mathrm{inf}}) & \approx & \frac{3}{8} \sqrt{3 F}. 
\end{eqnarray}
Any straight line, which has a slope greater than the maximum slope, 
intersects the Airy function at only one point. If the slope of the 
line is less than the maximum slope the line can intersect the 
Airy function at more than one point. 

The integral of Airy function can be found in terms of elementary functions 
\begin{equation}
   \int\limits_0^{\phi} A(\phi') d \phi' = 
   \frac{\arctan ( \sqrt{F + 1} \tan \phi )}{\sqrt{F + 1}}.
\end{equation}
This expression is used in defining the potential for the 
radiation pressure in the adiabatic approximation.


\newpage
\begin{thebibliography}{10}

\bibitem{Bradaschia:1990}
C.~Bradaschia, R.~Delfabbro, A.~Divirgilio, A.~Giazotto, H.~Kautzky,
  V.~Montelatici, D.~Passuello, A.~Brillet, and O.~Cregut et~al.
\newblock Terrestrial gravitational noise on a gravitational wave antenna.
\newblock {\em Nuclear Instruments A}, 289:518, 1990.

\bibitem{Abramovici:1992}
A.~Abramovici, W.E. Althouse, R.W. Drever, Y.Gursel, S.~Kawamura, F.J. Raab,
  D.~Shoemaker, L.~Sievers, R.E. Spero, K.~Thorne, R.E. Vogt, R.~Weiss, S.E.
  Whitcomb, and E.~Zucker.
\newblock {LIGO}: The {L}aser {I}nterferometer {G}ravitational-wave
  {O}bservatory.
\newblock {\em Science}, 256:281--412, April 17 1992.

\bibitem{Tsubono:1995}
K.~Tsubono.
\newblock 300-m laser interferometer gravitational wave detector ({TAMA} 300)
  in {J}apan.
\newblock In {\em Proceedings of First Eduardo Amaldi conference on
  gravitational wave experiments, Frascati, Roma, June 1994}, pages 112--114,
  Singapore, 1995. World Scientific.

\bibitem{Minorsky:1962}
N.~Minorsky.
\newblock {\em Nonlinear Oscillations}.
\newblock D. Van Nostrand Company, Inc., 1962.

\bibitem{Braginskii:1977}
V.B. Braginskii and A.B. Manukin.
\newblock {\em Measurement of weak forces in physics experiments}.
\newblock University of Chicago Press, 1977.

\bibitem{Dorsel:1983}
A.~Dorsel et~al.
\newblock Optical bistability and mirror confinement induced by radiation
  pressure.
\newblock {\em Physical Review Letters}, 51(17):1550--1553, October 1983.

\bibitem{Dorsel:1984}
A.~Dorsel et~al.
\newblock Optical resonators driven by radiation pressure.
\newblock {\em Philosophical Transactions of the Royal Society of London A},
  313:341--347, 1984.

\bibitem{Dorsel:1985}
A.~Dorsel et~al.
\newblock Light-pressure mirror stabilization.
\newblock {\em Acta Physica Austriaca}, 57:133--138, 1985.

\bibitem{Meystre:1985}
P.~Meystre et~al.
\newblock Theory of radiation-pressure-driven interferometers.
\newblock {\em Journal of the Optical Society of America B}, 2(11):1830--1839,
  November 1985.

\bibitem{Deruelle:1984}
N.~Deruelle and P.~Tourrenc.
\newblock The problem of the optical stability of a pendular {F}abry-{P}erot.
\newblock In {\em Gravitation, Geometry and Relativistic Physics}, pages
  232--237, Berlin, 1984. Springer-Verlag.

\bibitem{Tourrenc:1985}
Ph. Tourrenc and N.~Deruelle.
\newblock Effects of the time delays in a non linear pendular {F}abry-{P}erot.
\newblock {\em Annales de Physique}, 10:241--252, June 1985.

\bibitem{Aguirregabiria:1987}
J.~Aguirregabiria et~al.
\newblock Delay-induced instability in a pendular {F}abry-{P}erot cavity.
\newblock {\em Physical Review A}, 36(8):3768--3770, October 15 1987.

\bibitem{Bel:1988}
L.~Bel et~al.
\newblock Pendular {F}abry-{P}erot cavities as a paradigm for the dynamics of
  system with delays.
\newblock {\em Physical Review A}, 37(5):1563--1570, March 1988.

\bibitem{Meers:1989}
B.~Meers and N.~MacDonald.
\newblock Potential radiation-pressure-induced instabilities in cavity
  interferometers.
\newblock {\em Physical Review A}, 40(7):3754--3763, October 1989.

\bibitem{Chickarmane:1998}
V.~Chickarmane, S.V. Dhurandhar, R.~Barillet, P.~Hello, and J.-Y. Vinet.
\newblock Radiation pressure and stability of interferometric
  gravitational-wave detectors.
\newblock {\em Applied Optics}, 37(15):3236--3245, May 1998.

\bibitem{Rakhmanov:1998b}
M.~Rakhmanov.
\newblock Dynamics of {F}abry-{P}erot resonators with suspended mirrors. 2.
  {D}elay effects and control system.
\newblock {LIGO} technical report {T}970230, California Institute of
  Technology, February 1998.

\bibitem{Born:1980}
M.~Born and E.~Wolf.
\newblock {\em Principles of Optics}.
\newblock Pergamon Press, Oxford, 6 edition, 1980.

\bibitem{Marburger:1978}
J.H. Marburger and F.S. Felber.
\newblock Theory of a lossless nonlinear {F}abry-{P}erot interferometer.
\newblock {\em Physical Review A}, 17(1):335--342, January 1978.

\bibitem{Rowlands:1990}
G.~Rowlands.
\newblock {\em Non-linear phenomena in science and engineering}.
\newblock Ellis Horwood, 1990.

\bibitem{Andronov:1966}
A.A. Andronov, A.A. Vitt, and S.E. Khaikin.
\newblock {\em Theory of Oscillators}.
\newblock Pergamon Press, Oxford, New York, 1966.

\bibitem{Gibbs:1985}
H.M. Gibbs.
\newblock {\em Optical Bistability: Controlling Light with Light}.
\newblock Academic Press, Inc., Orlando, Florida, 1985.

\bibitem{Wherrett:1985}
B.S. Wherrett and S.D.~Smith F.R.S, editors.
\newblock {\em Optical Bistability, Dynamical Nonlinearity and Photonic Logic},
  London, 1985. A Royal Society Discussion Meeting held on 21 and 22 March
  1984, organized by S.D. Smith, F.R.S, A. Miller and B.S. Wherrett, The Royal
  Society.

\bibitem{Arnold:1984}
V.I. Arnold.
\newblock {\em Catastrophe theory}.
\newblock Springer-Verlag, Berlin, New York, 1984.

\bibitem{Landau:Mech}
L.D. Landau and E.M. Lifshitz.
\newblock {\em Mechanics}.
\newblock Pergamon Press, Oxford, New York, 3 edition, 1989.

\end{thebibliography}
\end{document}